\RequirePackage{fix-cm}
\documentclass[smallcondensed]{svjour3}     
\smartqed  
\usepackage{graphicx}
\usepackage{color}
\usepackage{epsfig}
\usepackage{amssymb}
\usepackage{amsmath}
\usepackage{bm}
\usepackage{algorithm}
\usepackage{algorithmic}

\usepackage{color}
\usepackage{epsfig}
\usepackage{amssymb}
\usepackage{amsmath}
\usepackage{amsfonts}
\usepackage[normalem]{ulem}

\usepackage{mwe}
\usepackage{graphbox}

\usepackage{array}

\newcommand{\bR}{{\mathbb R}}

\begin{document}

\title{Estimation and uncertainty quantification for piecewise smooth signal recovery
\thanks{This work was supported in part by NSF-DMS 1502640, NSF-DMS 1912685, and AFOSR FA9550-18-1-0316.}
}

\titlerunning{High order total variation Bayesian learning}        

\author{Victor Churchill         \and
        Anne Gelb
}


\institute{V. Churchill \at
              Department of Mathematics, Dartmouth College, Hanover, NH, USA \\
              \email{Victor.A.Churchill.GR@dartmouth.edu}           
           \and
           A. Gelb \at
              Department of Mathematics, Dartmouth College, Hanover, NH, USA\\
              \email{Anne.E.Gelb@dartmouth.edu}           
}

\date{Received: date / Accepted: date}

\maketitle

\begin{abstract}
This paper presents a sparse Bayesian learning (SBL) algorithm for linear inverse problems with a high order total variation (HOTV) sparsity prior. For the problem of sparse signal recovery, SBL often produces more accurate estimates than maximum \emph{a posteriori} estimates, including those that rely on $\ell_1$ regularization. Moreover, rather than a single signal estimate, SBL yields a full posterior density estimate which can be used for uncertainty quantification. However, SBL is only immediately applicable to problems having a {\em direct} sparsity prior, or to those that can be formed via synthesis. This paper demonstrates how a problem with an HOTV sparsity prior can be formulated via synthesis, and then develops a corresponding Bayesian learning method. This expands the class of problems available to Bayesian learning to include, e.g., inverse problems dealing with the recovery of piecewise smooth functions or signals from data. Numerical examples are provided to demonstrate how this new technique is effectively employed.
\end{abstract}

\section{Introduction}
\label{sec:intro}

Many real-world phenomena give rise to piecewise smooth signals, \cite{mallat1999wavelet}. As such, inverse problems to recover them from measurement data is a well-studied problem, \cite{stefan2010improved}. Particular attention has been paid to piecewise smooth signal or function recovery from Fourier or spectral data, \cite{gelb2000hybrid,gelb2007reconstruction,gelb2002spectral}. A now standard approach to piecewise smooth signal recovery is to minimize a least squares cost function with $\ell_1$ norm based high order total variation (HOTV) regularization, \cite{chan2000high,rudin1992nonlinear}. This is well-known to encourage sparsity in the approximate edge domain of the function. In practice, this can be achieved by penalizing the gradient domain of the signal using the HOTV operator ${\bm T}_m\in\mathbb{R}^{(N-m)\times N}$, a finite difference approximation to the $m$th gradient. In particular, this paper deals with HOTV orders $m=1,2,3$. While we do not explicitly consider $m\ge4$, the methods developed are easily adapted. Using such an operator is common for inverse problems in image processing when one has a prior belief that the signal of interest being recovered is approximately piecewise polynomial of order $m-1$, \cite{archibald2016image}. This technique has been useful in applications to improve robustness in synthetic aperture radar imaging, \cite{archibald2016image,sanders2017composite}, and to recover fine details in electron tomography imaging, \cite{sanders2017recovering}.

The main contribution of this paper is an alternative Bayesian learning based method for inverse problems where an HOTV sparsity prior is appropriate. This expands the class of problems available to this clearly strong method which provides a full posterior density estimate rather than a single point estimate. Because the sparsity assumption for piecewise smooth signal reconstruction is typically viewed in the analysis formulation, i.e. ${\bm T}_m{\bm x}={\bm s}$ with $\bm{x}$ the signal of interest and $\bm{s}$ the sparse representation, Bayesian learning is not immediately applicable. In particular, since ${\bm T}_m$ is not square and therefore not invertible, more work is required. In what follows, our approach is to form an equivalent synthesis formulation of the form $\bm{x} = \bm{Vs}$ in order to effectively reduce the problem to sparse signal recovery. Since sparse Bayesian learning (SBL), \cite{tipping2001sparse}, is applicable and has been shown to be superior to many other methods for sparse signal recovery, \cite{giri2016type,ji2008bayesian}, then one can expect superiority in this synthesis construction as well. Our procedure involves a modification from \cite{ortelli2019synthesis} to the analysis operators $\bm{T}_m$ to make these operators full rank and therefore invertible. This ultimately enables the formulation of a Bayesian learning algorithm for inverse problems with a HOTV sparsity prior like piecewise smooth signal recovery.

This paper is organized as follows. Section \ref{sec:background} reviews sparse signal recovery using a maximum a posteriori estimate, and describes how both the synthesis and analysis approaches are typically employed to recover signals that are sparse in a transform domain (e.g.~the HOTV domain). Section \ref{sec:synthesis} explains how to formulate a synthesis approach for the HOTV analysis operators via the technique introduced in \cite{ortelli2019synthesis}. Since SBL typically provides superior performance for the synthesis approach, in Section \ref{sec:estimation} we demonstrate how SBL specifically can be applied to synthetic HOTV. Numerical examples are implemented in Section \ref{sec:results}, where we demonstrate that our new approach, which we call high order total variation Bayesian learning (HOTVBL), outperforms the standard $\ell_1$ norm based HOTV regularization (analysis approach). Some concluding remarks and ideas for future investigations are provided in Section \ref{sec:conclusion}.

\section{Background}\label{sec:background}

\subsection{Sparse signal recovery}\label{subsec:ssr}
Let ${\bm x} \in \bR^N$  be a sparse signal with $k \ll N$ of its elements nonzero.   We seek to recover ${\bm x}$ from measurements 
\begin{equation}
\label{eq:measurements}
{\bm  b} = {\bm A}{\bm x}+{\bm n},
\end{equation}
where ${\bm A}\in\mathbb{R}^{J\times N}$ is a given forward measurement matrix and the given data is ${\bm b}\in\mathbb{R}^J$. The vector ${\bm n}\in\mathbb{R}^J$ is a noise vector accounting for model and measurement error. In this paper we will assume that $\bm{n}$ is zero-mean white Gaussian with variance $\nu^2$, and under the assumption that the entries of $\bm{b}$ are independent as in \cite{ji2008bayesian}, we have the likelihood model
\begin{equation}
\label{eq:glm}
p({\bm b}|{\bm x}) = (2\pi\nu^2)^{-J/2}\exp\left(-\frac{1}{2\nu^2}||{\bm A}{\bm x}-{\bm b}||_2^2\right).
\end{equation}
A straightforward way to estimate ${\bm x}$ is to maximize this likelihood:
\begin{eqnarray}
\label{eq:ssr}
{\bm x}^*_{ML} &=& \arg\max_{\bm x} \left\{ p({\bm b}|{\bm x})\right\}\nonumber\\
&=& \arg\max_{\bm x}\left\{ (2\pi\nu^2)^{-J/2}\exp\left(-\frac{1}{2\nu^2}||{\bm A}{\bm x}-{\bm b}||_2^2\right) \right\} \nonumber\\
&=& \arg\min_{\bm x} \left\{ ||{\bm A}{\bm x}-{\bm b}||_2^2\right\}.
\end{eqnarray}
However, solving (\ref{eq:ssr}) frequently yields a solution that is not sparse, i.e. with many greater than $k$ nonzero elements. To see this, consider the denoising problem where $J = N$ and $\bm{A}$ is the identity matrix. In this case the estimate is just the noisy collected signal, $\bm{x}^*_{ML} \equiv \bm{b}$. To improve on this result, a prior on $\bm{x}$ is often incorporated to encourage sparsity. For example, the Laplace density function
\begin{equation}\label{eq:laplaceprior}
p(\bm{x}) = \left(\frac{\mu}{2}\right)^{N}\exp\left(-\mu||\bm{x}||_1\right),
\end{equation}
is frequently chosen because it corresponds to the $\ell_1$ regularization often used in compressive sensing. Here $\mu$ determines the spread of the distribution and can intuitively be associated with how sparse $\bm{x}$ is. We note, however, that there are many sparsity-encouraging priors characterized by sharp peaks at zero. E.g., $||\bm{x}||_1$ in \eqref{eq:laplaceprior} can also be replaced with $||\bm{x}||_p^p$ for $p\in(0,1]$ which would correspond to $\ell_p$ regularization. Using Bayes' theorem we can now compute a maximum \emph{a posteriori} (MAP) estimate by maximizing the posterior
\begin{eqnarray}
{\bm x}^*_{MAP} &=& \arg\max_{\bm x} \left \{ p({\bm x}|{\bm b}) \right \} \nonumber\\
&=& \arg\max_{\bm x} \left \{ \frac{p({\bm b}|{\bm x}) p({\bm x})}{p({\bm b})} \right \}\nonumber\\
&=& \arg\max_{\bm x} \left \{ -\log p({\bm b}|{\bm x})-\log p({\bm x}) \right\} \nonumber\\
&=& \arg\min_{\bm x} \left \{ ||{\bm A}{\bm x}-{\bm b}||_2^2+\mu\nu^2||{\bm x}||_1\right\}.
\label{eq:map}
\end{eqnarray}
The first term in (\ref{eq:map}), often referred to as the fidelity term, is minimized when the solution aligns most closely with the given data. The second term is the imposed sparsity penalty on $\bm{x}$. In the field of compressive sensing the sparsity prior parameter $\mu$ and noise variance $\nu^2$ are often combined as $\lambda = \mu\nu^2$ and relabeled as the regularization parameter, which balances the fidelity term, the sparsity penalty, and noise reduction. Even though the inversion can be ill-posed, if certain conditions are met, then with high probability $\bm{x}$ can be exactly recovered from many fewer than $N$ measurements using this method, \cite{candes2006robust}. From \eqref{eq:map} it is also evident that without prior information of $\mu$ or $\nu^2$, it can be difficult to choose a suitable regularization parameter for any given application.

Moreover, even if the prior parameters are known, the maximum is not categorically representative of the posterior density. Hence it may be favorable to estimate the entire posterior density and \emph{then} derive statistics. In SBL, \cite{tipping2001sparse}, a flexible, hierarchical prior whose parameters are learned from the data is used to encourage sparsity and estimate the entire posterior density. Confidence intervals can be derived from this posterior to aid in uncertainty quantification, \cite{ji2008bayesian}. For sparse signal recovery, in terms of accuracy at a given sparsity level $k$, SBL has outperformed a variety of other methods, \cite{giri2016type,ji2008bayesian}, including the $\ell_1$ regularization scheme in \eqref{eq:map}, \cite{tibshirani1996regression}, and more advanced reweighting algorithms, \cite{candes2008enhancing,chartrand2008iteratively}. For more evidence see, e.g., Figs. 6, 7, and 8 in \cite{giri2016type}, and Figs. 2 and 4 in \cite{ji2008bayesian}.

\subsection{Synthesis and analysis}\label{subsec:synthesisandanalysis}

It is important to note that while $\ell_1$ and reweighted regularization schemes are readily adapted to signal processing applications, i.e. where the sparsity occurs in some related domain (e.g.~the gradient or wavelet domain), SBL is specifically designed for sparse signals.  Because of its clear advantages in sparse signal recovery, we would like to adapt SBL for these other problems as well.  Specifically, we are interested in applying SBL when the vector of interest ${\bm x}$ is not sparse, but a known transformation of ${\bm x}$ is. There are two approaches: synthesis and analysis.

In the \emph{synthesis} approach, which is typically associated with compressed sensing, we formulate a method based on the assumption that 
\begin{equation}
\bm{x} = \bm{V}\bm{s},
\label{eq:xVs}
\end{equation}
where $\bm{V}\in\mathbb{R}^{N\times M}$ is called a synthesis operator and $\bm{s}\in\mathbb{R}^{M}$ is a sparse vector. Sparse signal recovery is used to obtain $\bm{s}$ from measurements in (\ref{eq:measurements}), now written as  $\bm{b}=\bm{AVs}+\bm{n}$.   Using $\ell_1$ regularization as in (\ref{eq:map}), a synthesis approach recovery for the signal of interest $\bm{x}$ is given by
\begin{equation}\label{eq:l1synthesis}
{\bm x}^*_{synthesis} = {\bm V}\left[\arg\min_{\bm s} \left\{ ||{\bm A}{\bm V}{\bm s}-{\bm b}||_2^2+\lambda||{\bm s}||_1\right\}\right].
\end{equation}
Because synthesis effectively reduces the problem to sparse signal recovery, the SBL method is directly applicable to problems formed as (\ref{eq:l1synthesis}) by recovering ${\bm s}$ and then synthesizing via (\ref{eq:xVs}). In the corresponding \emph{analysis} approach, we formulate a method based on the assumption that 
\begin{equation}
{\bm T}{\bm x}={\bm s}, 
\label{eq:Txs}
\end{equation}
where ${\bm T}\in\mathbb{R}^{M\times N}$ is called an analysis operator and ${\bm s}\in\mathbb{R}^{M}$ is a sparse vector. In an $\ell_1$ regularization scheme, the signal of interest ${\bm x}$ is directly estimated by regularizing on the sparsity of ${\bm T}{\bm x}$ as
\begin{equation}\label{eq:l1analysis}
{\bm x}^*_{analysis} = \arg\min_{\bm x} \left\{||{\bm A}{\bm x}-{\bm b}||_2^2+\lambda||{\bm T}{\bm x}||_1\right\}.
\end{equation}
Choosing to use (\ref{eq:l1synthesis}) or (\ref{eq:l1analysis}) may simply depend on whether it is more natural to view the sparsifying transformation as ${\bm T}{\bm x}={\bm s}$ or ${\bm x} ={\bm V}{\bm s}$. The choice of analysis versus synthesis and the differences and similarities between them are further analyzed in \cite{elad2007analysis}. A particular result of interest is that if ${\bm V} := {\bm T}^{-1}$, then \eqref{eq:l1synthesis} and \eqref{eq:l1analysis} are equivalent. As discussed in more detail in Section \ref{sec:SBL}, because of the conjugate prior structure used in the SBL method, SBL is not readily applicable to problems viewed in the analysis approach.

In this paper we focus on a particular problem that is typically viewed in the analysis approach, that is when the underlying signal can be viewed as a piecewise smooth function. We will mainly consider the case for which the analysis operator ${\bm T}$ is the high order total variation (HOTV) operator ${\bm T}_m\in\mathbb{R}^{(N-m)\times N}$, a finite difference approximation to the $m$th gradient. Using such an analysis operator is used in inverse problems when one has a prior belief that the signal of interest being recovered is approximately piecewise polynomial of order $m-1$, \cite{archibald2016image}. In particular, we are interested in $m=1,2,3$. While we do not explicitly consider $m\ge4$ in this paper, we provide general formulae for these cases.

Our goal is to formulate a Bayesian learning method for piecewise smooth signal recovery, or more generally inverse problems with a HOTV sparsity prior. This will expand the class of problems available to Bayesian learning. However, as mentioned above, the HOTV problem is viewed in the analysis approach with ${\bm T}_m{\bm x}={\bm s}$ as in (\ref{eq:Txs}), and ${\bm T}_m$ is not square and therefore not invertible. Hence SBL is not immediately applicable. In Section \ref{sec:synthesis}, we demonstrate how to form an equivalent synthesis operator for HOTV in order to reduce the problem to sparse signal recovery,  after which we can directly apply SBL.  As a consequence, this approach should yield the same benefits as SBL does for standard sparse signal recovery. Our procedure involves a modification from \cite{ortelli2019synthesis} to make the analysis operators full rank and therefore invertible. This ultimately enables us to then formulate a Bayesian learning algorithm for inverse problems with a HOTV sparsity prior in Section \ref{sec:estimation}.

\section{Synthesis Operators for HOTV Regularization}
\label{sec:synthesis}

In part because of its edge-preserving properties, HOTV regularization is a common technique for inverse problems in image processing, \cite{archibald2016image,sanders2017composite,sanders2017recovering}.  The corresponding HOTV operator, ${\bm T}_m\in\mathbb{R}^{(N-m)\times N}$ in (\ref{eq:l1analysis}), is a scaled finite difference approximation of the $m$th gradient.\footnote{As mentioned, we only consider $m \le 3$, which sufficiently captures the signal variation in our examples.  Higher order gradients may be more suitable in other applications, or when resolution is insufficient, \cite{sanders2017recovering}.} For example, when $N=6$ we have
\begin{equation}
\small{
{\bm T}_1 = \begin{bmatrix} -1& 1& 0& 0& 0 & 0\\ 0& -1& 1 &0 &0&0\\ 0& 0& -1& 1& 0&0\\ 0& 0 & 0 & -1& 1&0\\0&0&0&0&-1&1\end{bmatrix};
{\bm T}_2 = \begin{bmatrix} 1& -2& 1 &0 &0&0\\ 0& 1& -2& 1& 0&0\\ 0& 0 &1 &-2& 1&0\\0&0&0&1&-2&1\end{bmatrix};
{\bm T}_3 = \begin{bmatrix} -1& 3& -3& 1& 0&0\\ 0& -1 &3 &-3& 1&0\\0&0&-1&3&-3&1\end{bmatrix}.
}
\end{equation}
Clearly ${\bm T}_1{\bm x} = {\bm s}$ is sparse whenever the underlying signal ${\bm x}$ is piecewise constant, since ${\bm T}_1$ is an exact transformation to the edge domain. High order gradients are useful when it is assumed that the smooth regions of the signal are better approximated by piecewise polynomials. 

It has been demonstrated that SBL is more effective for sparse signal recovery than many other algorithms, including $\ell_1$ regularization and many variants, \cite{giri2016type,ji2008bayesian}. In the synthesis approach, problems with transform sparsity priors are essentially reduced to sparse signal recovery. Hence, since SBL is more effective for sparse signal recovery, it may be advantageous to use SBL whenever an analysis approach can be replaced by a synthesis approach. In addition, Bayesian learning is able to estimate a posterior distribution for the signal as opposed to a single signal estimate, which can aid in uncertainty quantification. In what follows we demonstrate this idea. Specifically, we employ SBL to estimate the HOTV sparsity representation of the signal and subsequently synthesize the piecewise smooth signal of interest. A density estimate for both the sparse representation and the signal of interest are obtained.

Since SBL is available {\em only} to problems formed via synthesis, we must first find a corresponding synthesis operator for ${\bm T}_m$. A problem quickly arises in developing a synthesis approach for HOTV, however. Notably, ${\bm T}_m$ is not invertible (or square), so the required synthesis operator ${\bm V}$ such that ${\bm x} = {\bm V}{\bm s}$ is not immediately apparent. Hence as in \cite{ortelli2019synthesis}, we ``complete'' ${\bm T}_m$, which we will denote as $\tilde{\bm T}_m$,  by adding rows in its null space.  As in Appendix D of \cite{ortelli2019synthesis}, rows corresponding to the $0$ through $(m-1)$th forward difference coefficients are added. For example, when $N=6$ and $m=1,2,3$ we have
\begin{equation}\label{eq:TV}
\small{
\tilde{\bm T}_1 = \begin{bmatrix} 1 & 0 &0 &0& 0&0\\ -1& 1& 0& 0& 0&0\\ 0& -1& 1 &0 &0&0\\ 0& 0& -1& 1& 0&0\\ 0& 0 & 0 & -1& 1&0\\0&0&0&0&-1&1\end{bmatrix};
\tilde{\bm T}_2  = \begin{bmatrix} 1 & 0 &0 &0& 0&0\\ -1& 1& 0& 0& 0&0\\ 1& -2& 1 &0 &0&0\\ 0& 1& -2& 1& 0&0\\ 0& 0 &1 &-2& 1&0\\0&0&0&1&-2&1\end{bmatrix};
\tilde{\bm T}_3  = \begin{bmatrix} 1 & 0 &0 &0& 0&0\\ -1& 1& 0& 0& 0&0\\ 1& -2& 1 &0 &0&0\\ -1& 3& -3& 1& 0&0\\ 0& -1 &3 &-3& 1&0\\0&0&-1&3&-3&1\end{bmatrix}.
}
\end{equation}
Observe that $\tilde{\bm T}_m \in {\mathbb R}^{N\times N}$ has rank $N$ and yields a new sparse representation
\begin{align}
\tilde{\bm T}_m{\bm x}={\bm t} = \begin{bmatrix} r \\\bm{s}\end{bmatrix},
\end{align}
where $r\in\mathbb{R}^m$ and $\bm{s}\in\mathbb{R}^{N-m}$, thus ${\bm t} \in {\mathbb R}^N$. Moreover, the matrix completion is constructed in a sensible way since the points added to the sparse representation are simply a finite difference approximation to the derivative. For example, if the previous stencil contained three points, such as the case corresponding to ${\bm T}_2$,  to construct the completion matrix $\tilde{\bm T}_2$ the coefficients of two point centered differencing are used in the second row. Since we can not approximate a derivative with one grid point, we simply use $1$ in the first row. The process is similar for  generating general $\tilde{\bm T}_m$, with each of the top $m$ rows (except the first) having values corresponding to the coefficients of the $m$th finite difference derivative approximations.

The synthesis operators for HOTV analysis operators in \eqref{eq:TV} are subsequently defined by ${\bm V}_m := \tilde{\bm T}_m^{-1}$.  For example, for $N = 6$ and $m = 1,2,3$ we have
\begin{align}
{\bm V}_1 = \begin{bmatrix} 1 & 0 &0 &0& 0&0\\ 1& 1& 0& 0& 0&0\\ 1& 1& 1 &0 &0&0\\ 1& 1& 1& 1& 0&0\\ 1& 1 &1 &1& 1&0\\1&1&1&1&1&1\end{bmatrix},
{\bm V}_2 = \begin{bmatrix} 1 & 0 &0 &0& 0&0\\ 1& 1& 0& 0& 0&0\\ 1& 2& 1 &0 &0&0\\ 1& 3& 2& 1& 0&0\\ 1& 4 &3 &2& 1&0\\1&5&4&3&2&1\end{bmatrix},
{\bm V}_3 = \begin{bmatrix} 1 & 0 &0 &0& 0&0\\ 1& 1& 0& 0& 0&0\\ 1& 2& 1 &0 &0&0\\ 1& 3& 3& 1& 0&0\\ 1& 4 &6 &3& 1&0\\1&5&10&6&3&1\end{bmatrix}.
\end{align}
In general ${\bm V}_m$ is lower triangular. A general formula for $\bm{V}_m$ for $m\ge2$ in terms of $\bm{V}_{m-1}$ is
\begin{align}
\bm{V}_m[i,j] = \left\{ \begin{matrix} \bm{V}_{m-1}[i,j] & \mbox{if $j<m$} \\ \sum_{k=1}^i \bm{V}_{m-1}[k,j] & \mbox{if $j\ge m$} \end{matrix} \right. .
\end{align}

The main result of \cite{ortelli2019synthesis} of use in this paper is Lemma 3.2, which asserts that assuming there is sparsity in the latter $N-m$ elements of the new sparse vector (i.e. the original sparse representation), the $\ell_1$ regularized estimate using $\tilde{\bm T}_m$ is consistent with the original problem using ${\bm T}$ since the added rows are in the null space of $\bm{T}$. In particular, the following two estimates via analysis and synthesis are shown to be equivalent:
\begin{align}\label{eq:finall1}
\begin{split}
\bm{x}^*_{\ell_1} &= \arg\min_{\bm x} \left\{||\bm{Ax}-\bm{b}||_2^2+\lambda||\bm{T}_m\bm{x}||_1\right\}\\
&= \bm{V}_m\left[\arg\min_{\bm t}\left\{ ||\bm{AV}_m\bm{t}-\bm{b}||_2^2+\lambda||\bm{t}[m+1:N]||_1\right\}\right],
\end{split}
\end{align}
where $\bm{t}[m+1:N]$ denotes the latter $N-m$ elements of $\bm{t}$. That is, in the synthesis form the sparsity-encouraging $\ell_1$ norm only regularizes with respect to the elements of the original analysis operation. We use this equivalency in the next section to inspire the use of $\bm{V}_m$ as a synthesis operator in a Bayesian learning procedure. In addition, our approach does not require the sparsity in the remaining $N-m$ elements of the new sparse vector (as assumed in Lemma 3.2 in  \cite{ortelli2019synthesis}), since the {\em data} directly dictate which elements in the sparsity domain have non-zero value.

\section{High Order Total Variation Bayesian Learning (HOTVBL)}\label{sec:estimation}

The MAP estimate provided in \eqref{eq:map} is typically aligned with the compressive sensing approach for sparse signal recovery, and forms the basis for the approximation in \eqref{eq:l1analysis} when the signal is sparse in some transform domain. As noted previously, the MAP estimate is not categorically representative of the posterior density. Because of this limitation, a better approach is needed.

In Bayesian learning, instead of a fixed sparsity-inducing prior on ${\bm t}= \bm{\tilde{T}}_m{\bm x}$, an empirical prior characterized by flexible parameters that must be estimated from the data is used. In this investigation we focus on sparse Bayesian learning (SBL), \cite{tipping2001sparse}, which has also been used in Bayesian compressed sensing, \cite{ji2008bayesian}. It is important to note that SBL is only available to problems formed via synthesis or directly sparse problems. Indeed, this is what motivated our derivation of the HOTV synthesis operator in Section \ref{sec:synthesis}. 

Recall that we seek to employ SBL since in many cases it has been shown empirically and theoretically to be superior in terms of accuracy to MAP estimates, \cite{faul2002analysis,giri2016type,wipf2004sparse,wipf2005norm}. Theoretical analysis in \cite{rao2006comparing} and \cite{wipf2005norm} shows that SBL provides a closer approximation to the $\ell_0$ norm of the sparse signal than the $\ell_1$ norm. For the noiseless case, it was shown in \cite{wipf2004sparse} that the global minimum of the effective SBL cost function is achieved at a solution such that the posterior mean equals the maximally sparse solution. Furthermore, local minima are achieved at sparse solutions, regardless of noise. Empirically, \cite{giri2016type} shows that SBL achieves superior sparse signal recovery results compared to $\ell_1$, reweighted $\ell_1$, and reweighted $\ell_2$ regularization (see \cite{candes2006robust,candes2008enhancing,chartrand2008iteratively}, respectively). This is further supported by multi-run testing in \cite{ji2008bayesian}. In addition, SBL provides a full posterior distribution and confidence intervals versus a point estimate, and automatically estimates all parameters from the given data.

Hence SBL will be used in an attempt to more accurately detect the sparse HOTV (or approximate edge) representation by recovering the sparse signal ${\bm t}=\bm{\tilde{T}}_m\bm{x}$ from noisy measurements
\begin{align}\label{eq:model}
{\bm b} &= \bm{Ax}+{\bm n} = \bm{AV}_m\bm{t}+{\bm n}:= \bm{H}_m\bm{t}+\bm{n},
\end{align}
where  ${\bm n}$ is distributed zero-mean Gaussian with unknown variance $\nu^2$.  The piecewise smooth signal is then recovered via synthesis by $\bm{x}={\bm V}_m\bm{t}$. 

\subsection{Sparse Bayesian Learning (SBL)}\label{sec:SBL}

Below is a brief review of how SBL is formulated.  More details can be found in \cite{ji2008bayesian,tipping2001sparse}. First we develop a parametrized prior on ${\bm t}$. Because Gaussian noise is assumed in (\ref{eq:model}), we define a conjugate zero-mean Gaussian prior on each element of ${\bm t}$
\begin{align*}
p({\bm t}|{\bm a}) = \prod_{i=1}^{N}\mathcal{N}({\bm t}_i|0,{\bm a}_i^{-1}),
\end{align*}
where ${\bm a}_i$ is the precision or inverse variance. We then define a conjugate Gamma prior over ${\bm a}$
\begin{align*}
p({\bm a}|a,b)=\prod_{i=1}^{N}\Gamma({\bm a}_i|a,b).
\end{align*}
Finally, we marginalize over the hyperparameters ${\bm a}$ to obtain the overall prior on ${\bm t}$ as
\begin{align}\label{eq:overallprior}
p({\bm t}|a,b) = \prod_{i=1}^{N}\int_0^\infty\mathcal{N}({\bm t}_i|0,{\bm a}_i^{-1})\Gamma({\bm a}_i|a,b)d{\bm a}_i.
\end{align}
Each integral being multiplied in (\ref{eq:overallprior}) is distributed via the Student's $t$-distribution, which, for suitable $a$ and $b$, is strongly peaked at ${\bm t}_i=0$. Therefore this prior favors ${\bm t}_i$ being zero, hence encouraging sparsity. We also impose a conjugate Gamma prior $\Gamma(\beta|c,d)$ on $\beta=\frac{1}{\nu^2}$. Only point estimates are needed for ${\bm a}$ and $\beta$, so we simply set $a,b,c,d=0$ implying uniform hyperpriors on a logarithmic scale for ${\bm a}$ and $\beta$, \cite{tipping2001sparse}. Because of the conjugate priors used above, the posterior distribution for ${\bm t}$ can be solved for analytically as a multivariate Gaussian distribution
\begin{align*}
p({\bm t}|{\bm b},{\bm a},\beta) &=\mathcal{N}({\bm t}|\boldsymbol{\mu},\boldsymbol{\Sigma}),
\end{align*}
with mean and covariance matrix given by
\begin{align}\label{eq:mean}
\boldsymbol{\mu} &= \beta\boldsymbol{\Sigma}\bm{H}_m^T{\bm b},
\end{align}
\begin{align}\label{eq:cov}
\boldsymbol{\Sigma} &= \left(\beta\bm{H}_m^T\bm{H}_m+{\bm \Lambda}\right)^{-1},
\end{align}
where ${\bm \Lambda}=\text{diag}({\bm a})$, \cite{bishop2006pattern}.

Marginalizing over ${\bm t}$, the marginal log-likelihood for ${\bm a}$ and $\beta$ is
\begin{align}\label{eq:L}
\begin{split}
\log p({\bm y}|{\bm a},\beta) &= \log\int p({\bm y}|{\bm g},\beta)p({\bm g}|{\bm a})d{\bm g}\\&=-\frac12\left(J\log 2\pi + \log |{\bm C}| +{\bm y}^t{\bm C}^{-1}{\bm y}\right),
\end{split}
\end{align}
with ${\bm C}=\beta^{-1}{\bm I}+\bm{H}_m {\bm \Lambda}^{-1}\bm{H}_m^T$, \cite{bishop2006pattern}.  Note that (\ref{eq:L}) cannot be maximized in closed form. In \cite{tipping2001sparse}, a maximum likelihood approximation is employed that uses the point estimates for ${\bm a}$ and $\beta$ to maximize (\ref{eq:L}), and is implemented via an {\em expectation-maximization} (EM) algorithm, \cite{dempster1977maximum}. In particular, the update for ${\bm a}$ to maximize (\ref{eq:L}) is
\begin{align}\label{eq:a}
{\bm a}_i^{\text{(new)}}=\frac{\gamma_i}{{\bm \mu}_i^2}
\end{align}
for each $i$, with ${\bm \mu}_i$ the $i$th posterior mean weight from (\ref{eq:mean}) and $\gamma_i = 1-{\bm a}_i\bm{\Sigma}_{ii}$ with $\bm{\Sigma}$ from (\ref{eq:cov}). For $\beta$ the update is
\begin{align}\label{eq:beta}
\beta^{\text{(new)}}=\frac{M-\sum_i\gamma_i}{||{\bm b}-\bm{H}_m{\bm \mu}||_2^2}.
\end{align}
Appendix A of \cite{tipping2001sparse} gives details on the derivation of these terms. Observe that ${\bm a}^{\text{(new)}}$ and $\beta^{\text{(new)}}$ are functions of ${\bm \mu}$ and $\boldsymbol{\Sigma}$, and vise versa. The EM algorithm iterates between (\ref{eq:mean}) and (\ref{eq:cov}), and (\ref{eq:a}) and (\ref{eq:beta}) until a convergence criterion is satisfied. Due to the properties of the EM algorithm, SBL is globally convergent, i.e. each iteration is guaranteed to reduce the cost function, \cite{wipf2004sparse}. It has been observed that most ${\bm a}_i\rightarrow\infty$, corresponding to a sparse result with many ${\bm t}_i\approx0$.

Note that after the convergence criterion has been satisfied, the final $\bm{\mu}^*$ and $\boldsymbol{\Sigma}^*$ are the mean and covariance matrix, respectively, of the Gaussian approximation to the posterior density function for ${\bm t}= \bm{\tilde{T}}_m{\bm x}$, \emph{not} ${\bm x}$. This density can perhaps be useful for tasks typically accomplished by edge detection such as boundary identification, scale separation, or other downstream processes such as determining the support of the signal, or which regions of the signal may need further investigation. While this approximate edge density may be of some use in and of itself, the approximate Gaussian posterior density for the piecewise smooth signal of interest ${\bm x}$ is defined by the statistics
\begin{eqnarray}\label{eq:finaldensity}
\text{E}({\bm x}) &= &{\bm V}_m{\bm \mu}^*,\nonumber\\
\text{Cov}({\bm x}) &=& {\bm V}_m\boldsymbol{\Sigma}^*{\bm V}_m^T.
\label{eq:EMalgorithm}
\end{eqnarray}
Similar to the case of sparse signal recovery, $\text{Cov}(\bm{x})$ can be used to develop confidence intervals for the estimated values of $\bm{x}$ as will be shown in Section \ref{sec:results}.

Note that each iteration of the described EM algorithm requires the inversion of an $N\times N$ matrix to compute the covariance matrix $\bm{\Sigma}$. This scales to $\mathcal{O}(N^3)$ operations -- clearly inefficient for large $N$. Fast algorithms based on the cost function (\ref{eq:L}) have been developed, \cite{faul2002analysis,tipping2003fast}, and are used in our numerical experiments. For signals of the size implemented in Section \ref{sec:results} (e.g., $N=128$ and $N=250$), we observed no difference between HOTVBL and the minimization of \eqref{eq:l1analysis} in terms of runtime.

\section{Numerical Results}\label{sec:results}

We now perform a variety of tests comparing HOTV $\ell_1$ estimates $\mathbf{x}^*_{\ell_1}$ given by \eqref{eq:finall1} and the HOTVBL procedure described in Section \ref{sec:SBL}. In particular we use $\mathbf{x}^*_{BL} = \text{E}(\mathbf{x})$ from \eqref{eq:finaldensity} as the point estimate associated with HOTVBL. The noise level in the collected data is measured by signal-to-noise ratio defined
\begin{align}
SNR = 20\cdot\log_{10}\left(\sqrt{\frac{\sum_{i=1}^N {\bm x}_i^2}{\sum_{i=1}^N{\bm n}_i^2}} \right).
\end{align}
We compare the reconstructions using the relative error defined
\begin{align}
RelErr({\bm x}^*) = \frac{||{\bm x}^* - {\bm x}||_2}{||{\bm x}||_2},
\end{align}
where ${\bm x}^*$ is the recovered signal and ${\bm x}$ is the ground truth. This provides a total error measure for each experiment. We also use the maximum error defined
\begin{align}
MaxErr(\bm{x}^*) = \arg\max_i |\bm{x}_i^* - \bm{x}_i|
\end{align}
to quantify the worst case pointwise error.

\begin{figure}[h!]
\centering
\includegraphics[width=.8\textwidth]{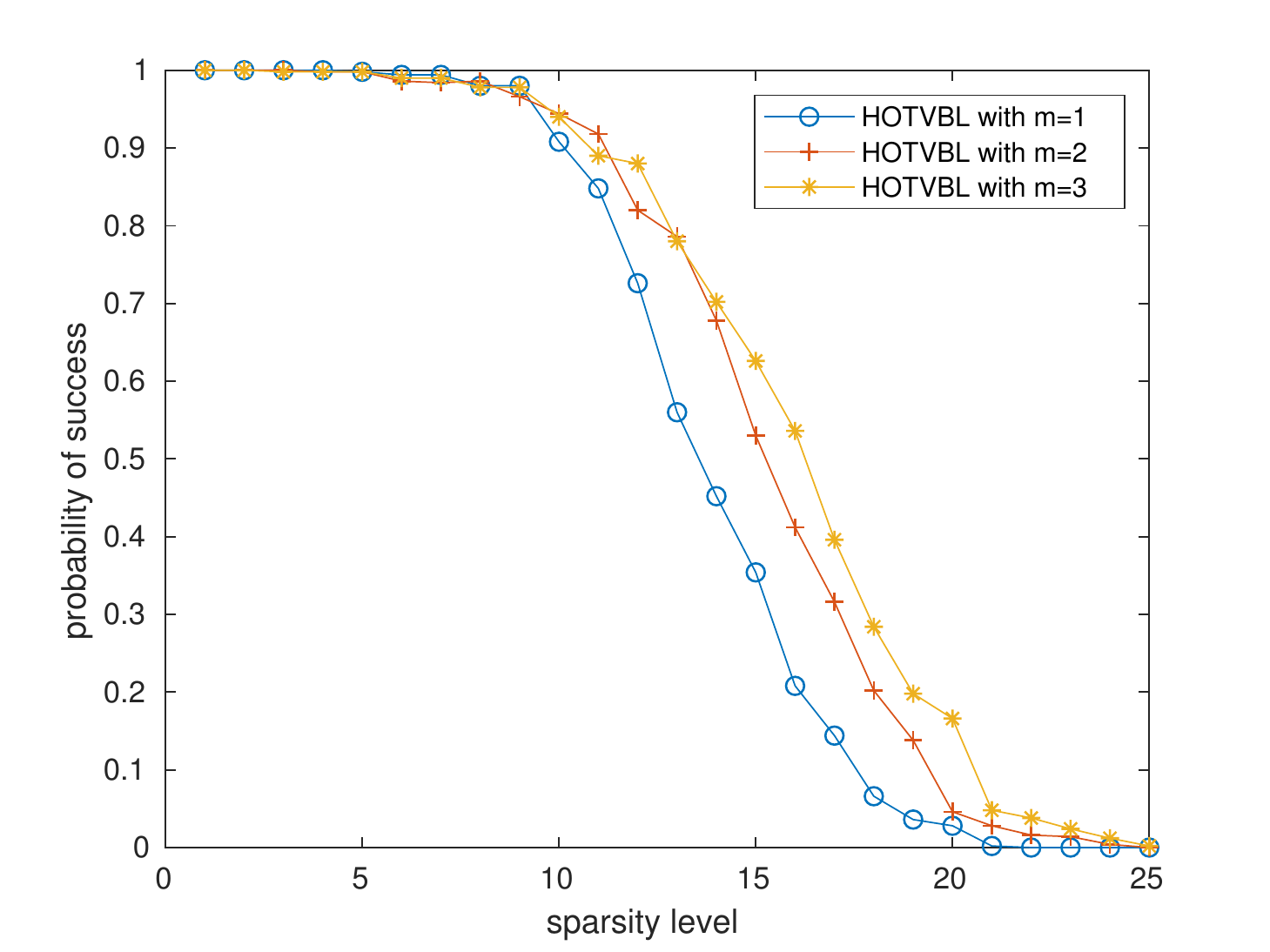}
\caption{Sparsity level versus probability of success.}
\label{fig:gaussiansuccessprobability}
\end{figure}

\paragraph{Test 1: Probability of success at a given sparsity level with underdetermined Gaussian forward model and no noise.} In this test, first a sparse signal $\bm{t}\in\mathbb{R}^{250}$ with $k$ nonzero elements is generated with the height of the spikes drawn from a standard Normal distribution. This signal is then transformed by $\bm{V}_m$ in order to generate a piecewise $(m-1)$ order polynomial $\bm{x}\in\mathbb{R}^{250}$ with $k$ jumps or edges whose heights are standard Normal distributed. This signal of interest is then obfuscated by a matrix $\bm{A}\in\mathbb{R}^{50\times250}$ where the entries of $\bm{A}$ have also been drawn from a standard Normal distribution. No noise is added, such that the data is modeled exactly by $\bm{b} = \bm{A}\bm{x}=\bm{H}_m\bm{t}$. The signal $\bm{x}$ is then reconstructed using HOTVBL with appropriate $m$. The above process represents one trial. A trial is considered a success if $MaxErr(\bm{x}^*_{BL})\le 10^{-3}$. For each $k=1,2,3,\ldots,25$, we ran $500$ trials, with the success probability determined as the number of successes divided by $500$. This is a standard test of sparse signal recovery methods, \cite{giri2016type}. Figure \ref{fig:gaussiansuccessprobability} shows the results for HOTVBL using $m=1,2,3$. Plots for the noise-free variant of \eqref{eq:finall1}, i.e.~basis pursuit \cite{chen2001atomic}, are omitted as there were \emph{no successes} registered in any of the 500 trials for any $k$ value. Note that this lack of success is likely due at least in part to the inherent regularization parameter $\lambda=1$ used in basis pursuit, \cite{boyd2011distributed}. In addition, the stringent success definition in terms of pointwise error likely ruled out many reconstructions whose relative error would be acceptable.

\paragraph{Test 2: Denoising reconstruction of ideal signals with varying noise level.} In this test, we consider the classical denoising problem, which epitomizes the difficulty in balancing fidelity, sparsity, and noise reduction. In denoising, ${\bm A}=\bm{I}$ the identity, meaning we collect a noisy signal ${\bm b} = {\bm x}+{\bm n}$, and regularize by the HOTV sparsity of the signal to return a result more faithful to the unknown ground truth signal. We compare the resulting reconstructions $\mathbf{x}^*_{\ell_1}$ from \eqref{eq:finall1} and the proposed HOTVBL procedure $\mathbf{x}^*_{BL}$. We test first on \emph{ideal} signals, that is ground truth signals that are exactly piecewise polynomial with only a single jump. In these cases, $\bm{\tilde{T}}_m$ is an appropriate sparsifying transform. Since the ground truth in this case is known, we can optimize the regularization parameter $\lambda$ in \eqref{eq:finall1} to minimize the relative error. We show this best-case scenario while noting that without oracle knowledge of the signal, this optimal result may be difficult to obtain in real-world examples.

\begin{figure}[h]
\centering
\includegraphics[width=.49\textwidth]{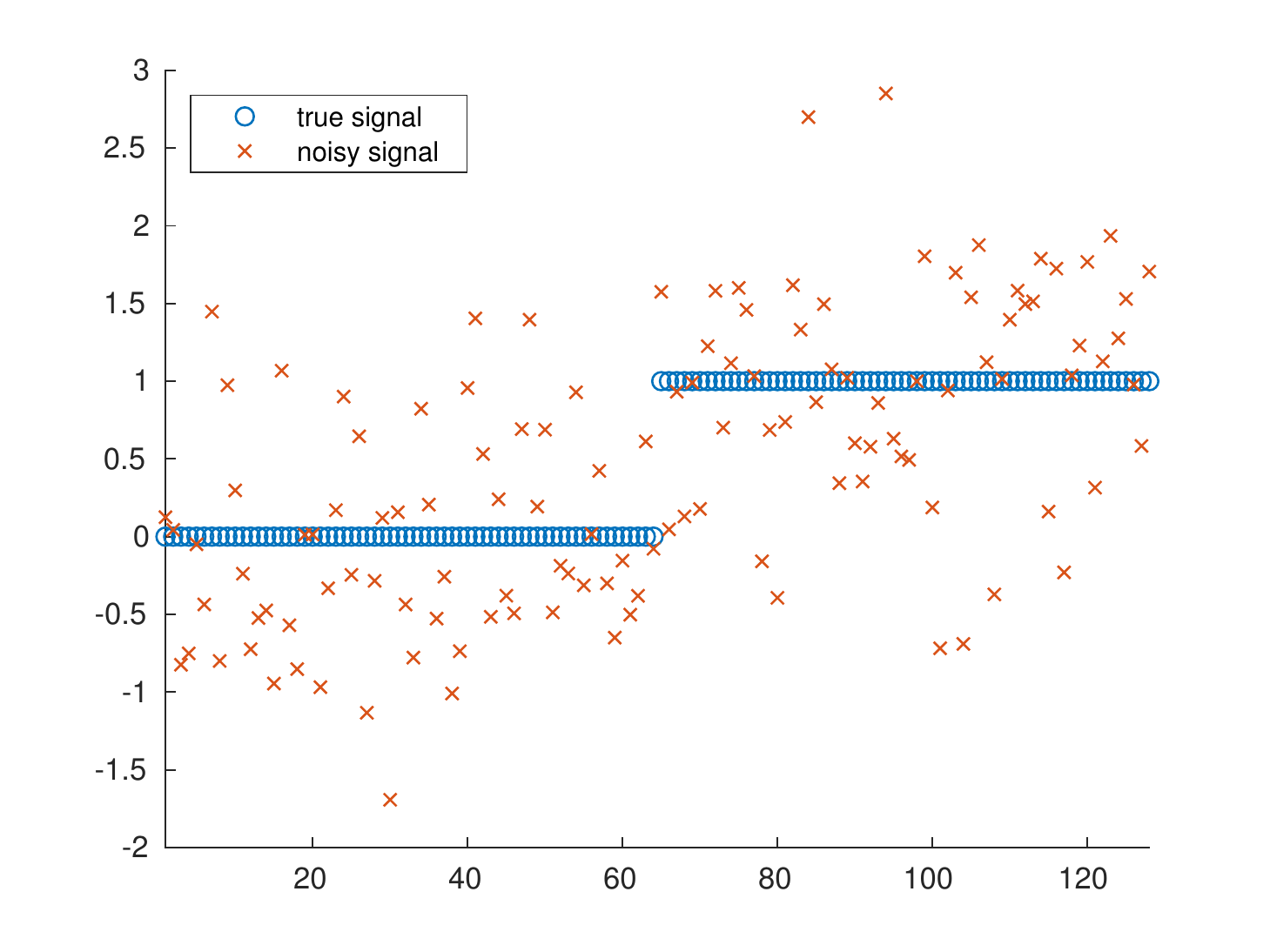}
\includegraphics[width=.49\textwidth]{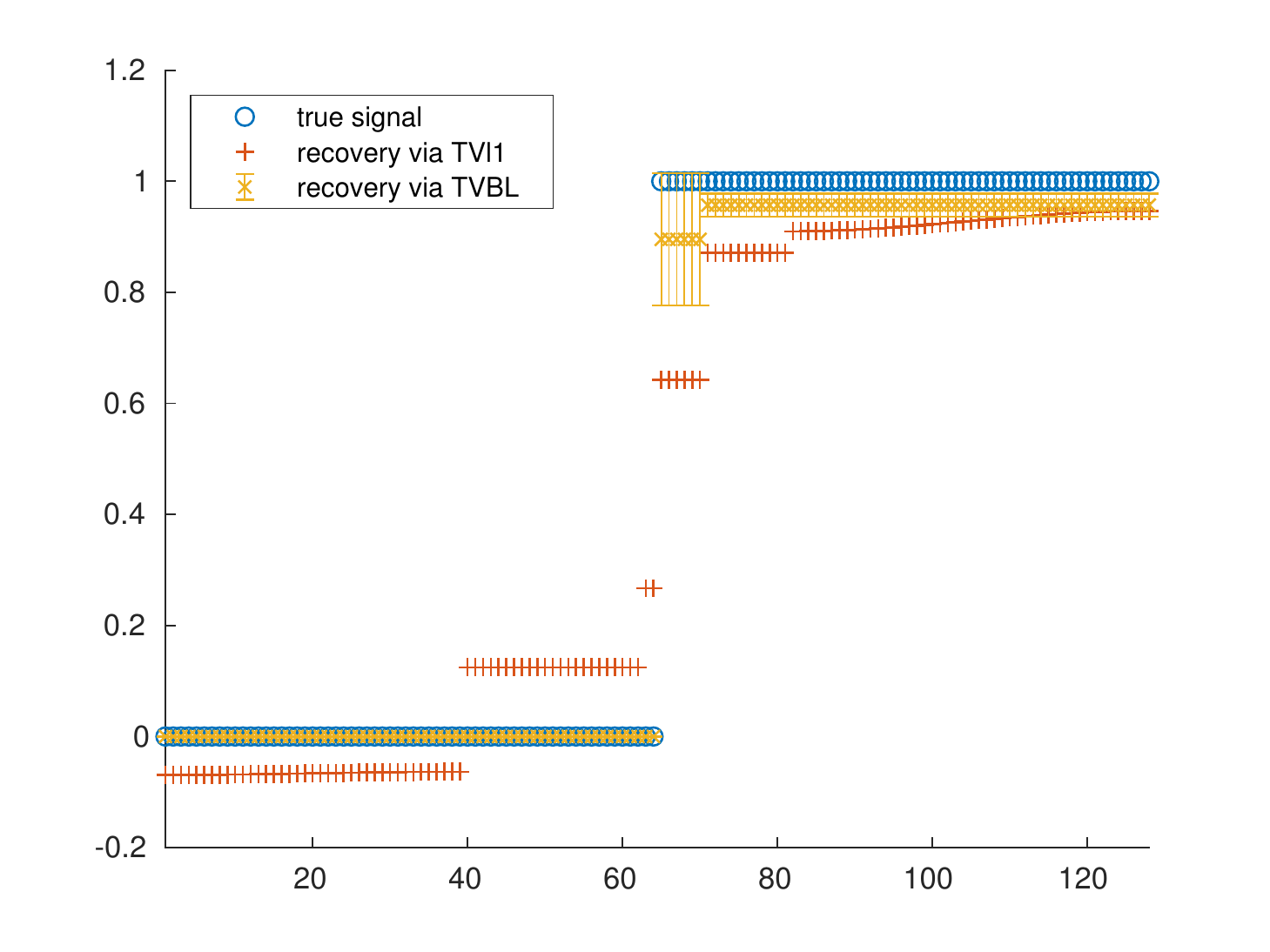}
\caption{Noisy, true, and recovered piecewise constant signals from noisy image data with SNR=0 dB using $m=1$.}
\label{fig:TVnoisy}
\end{figure}

\begin{figure}[h!]
\centering
\includegraphics[width=.49\textwidth]{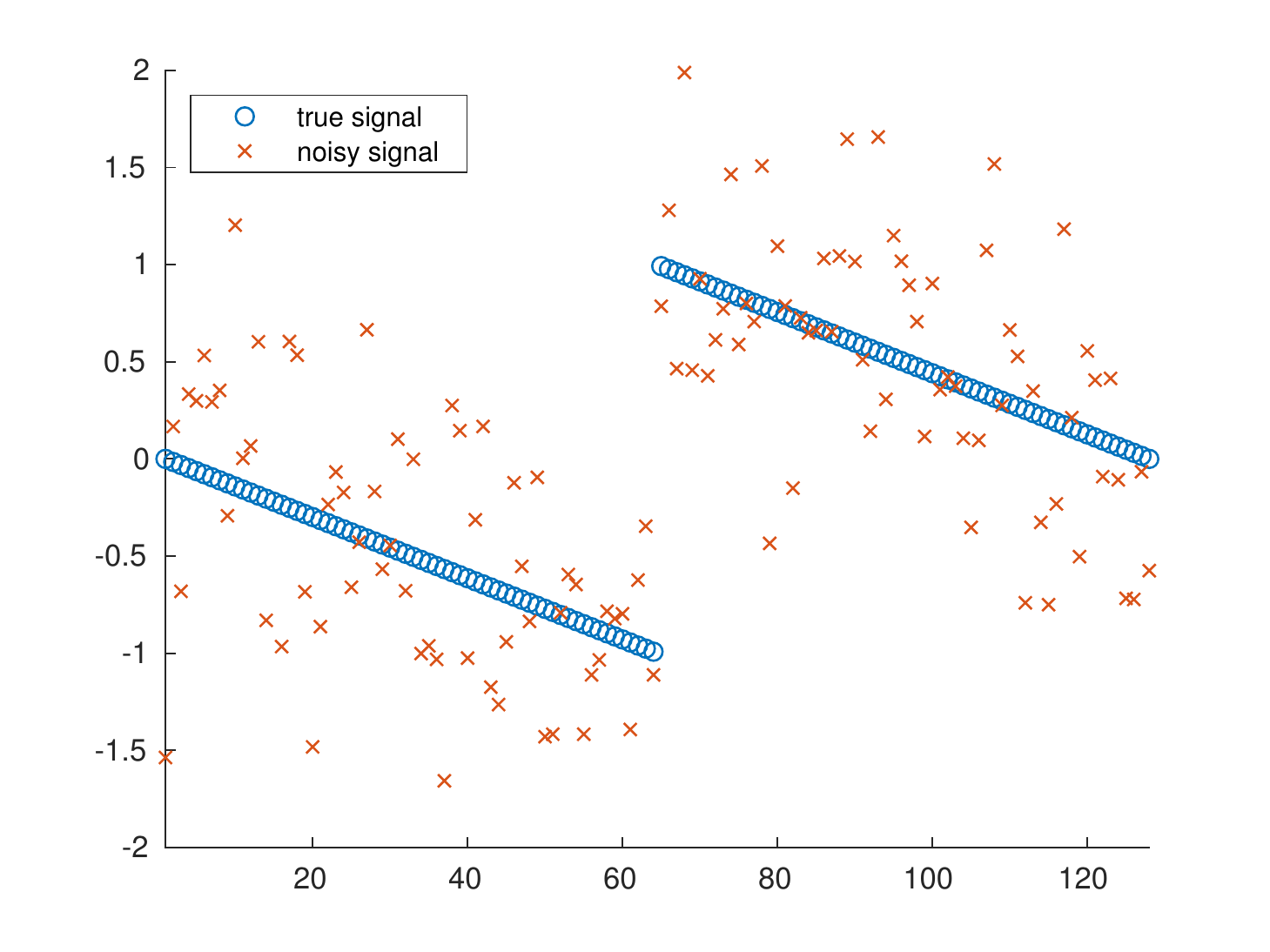}
\includegraphics[width=.49\textwidth]{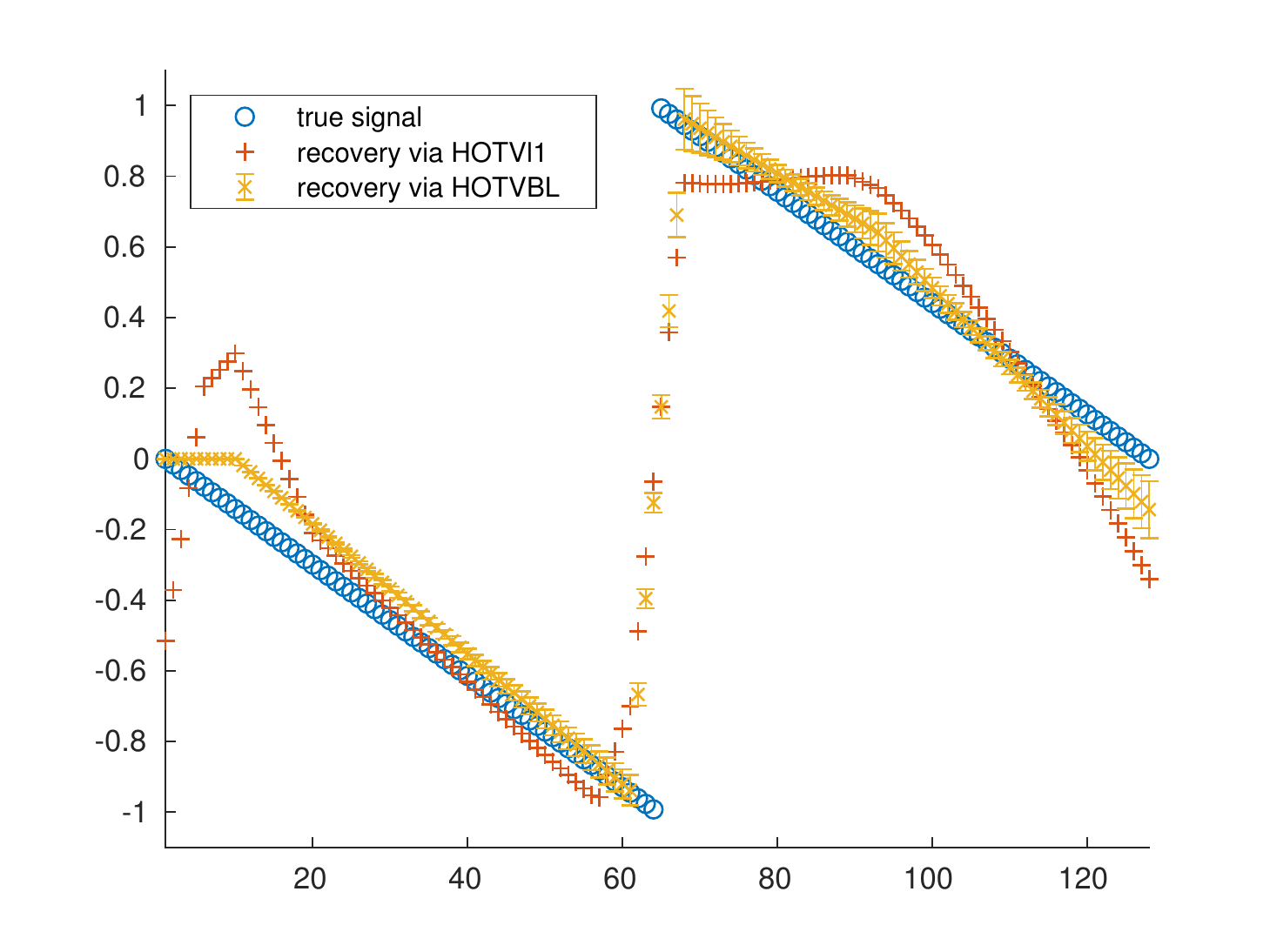}
\caption{Noisy, true, and recovered piecewise linear signals from noisy image data with SNR=0 dB using $m=2$.}
\label{fig:2TVnoisy}
\end{figure}

\begin{figure}[h!]
\centering
\includegraphics[width=.49\textwidth]{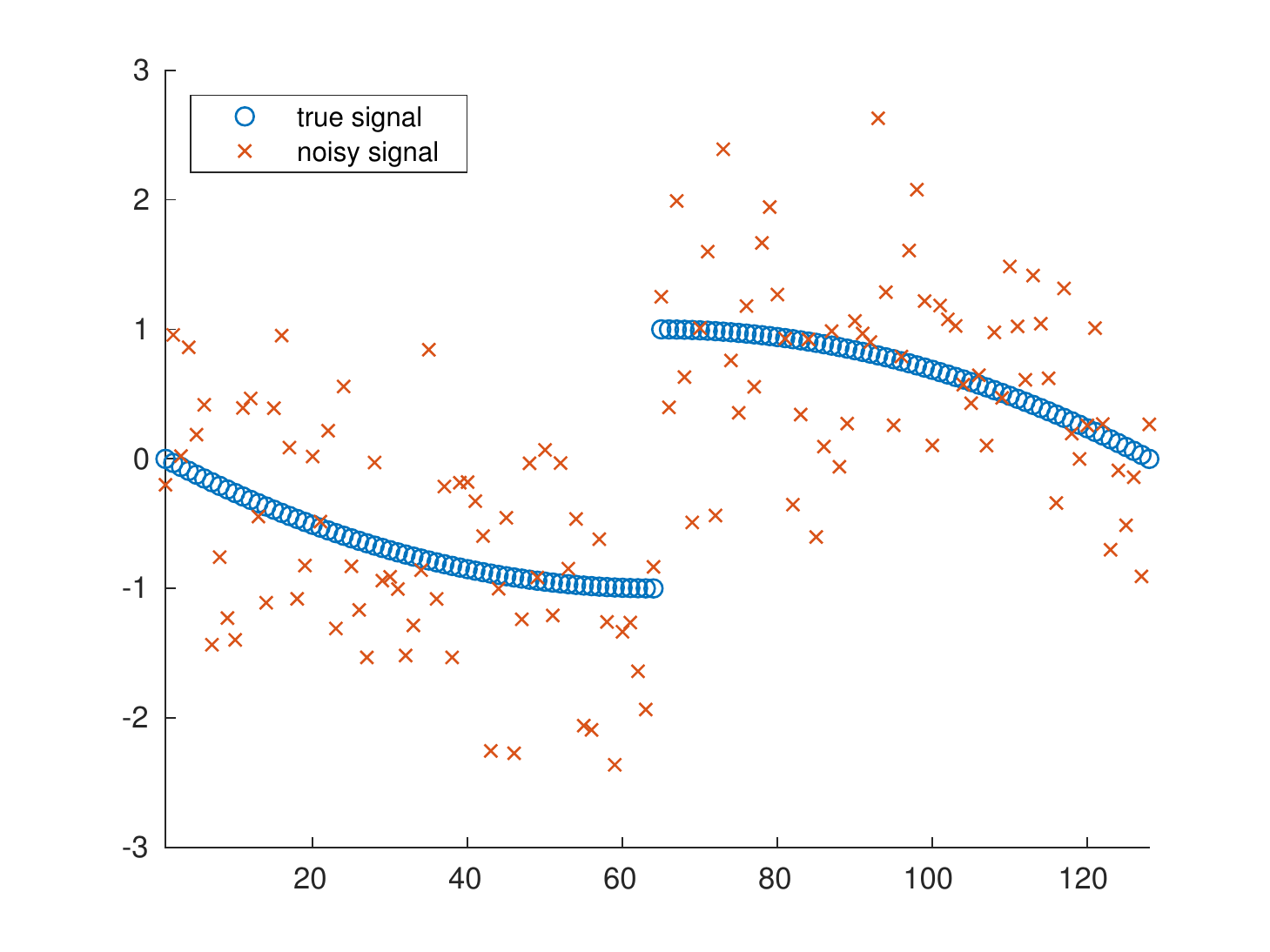}
\includegraphics[width=.49\textwidth]{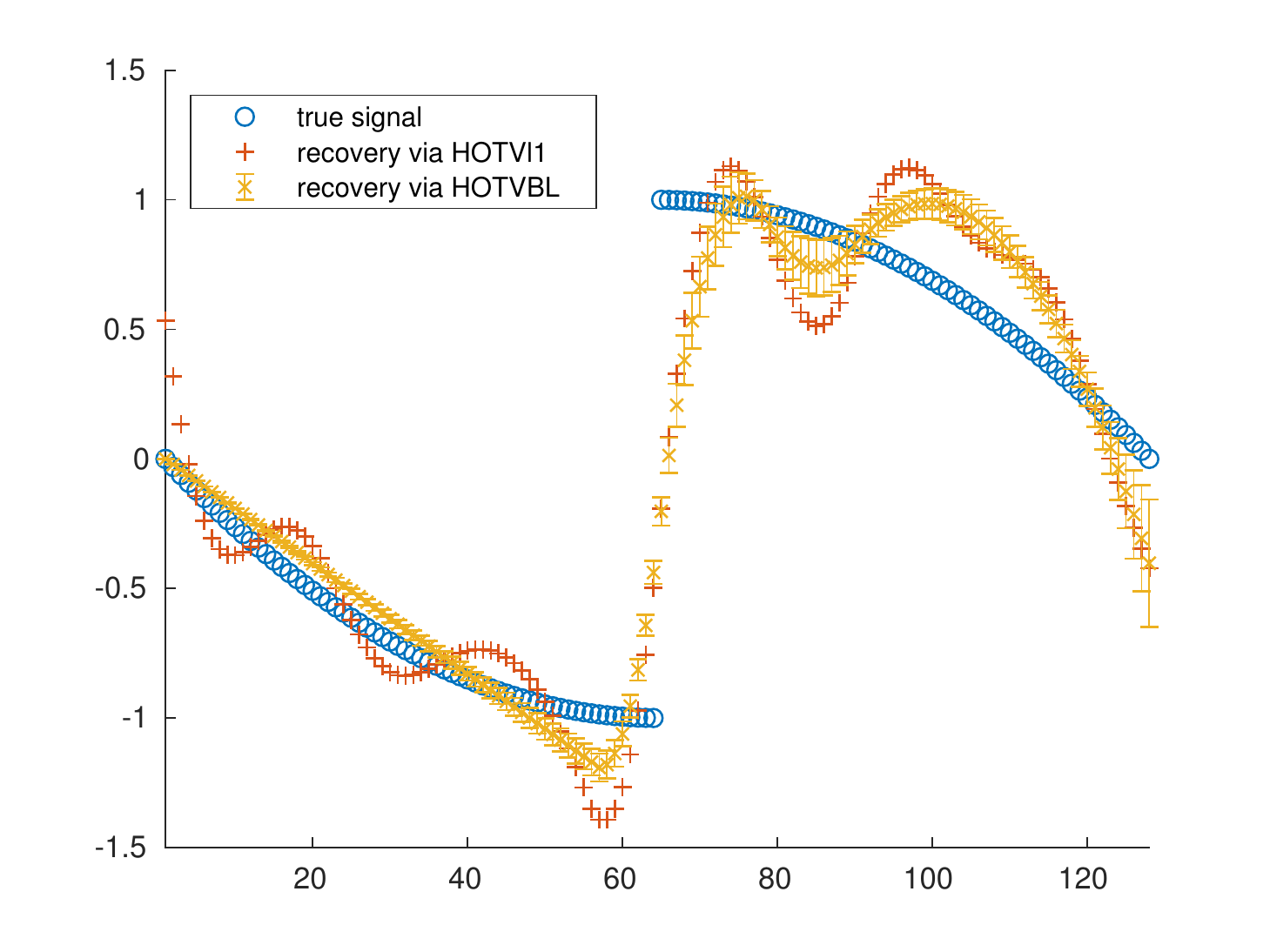}
\caption{Noisy, true, and recovered piecewise quadratic signals from noisy image data with SNR=0 dB using $m=3$.}
\label{fig:3TVnoisy}
\end{figure}

Figures \ref{fig:TVnoisy}, \ref{fig:2TVnoisy}, and \ref{fig:3TVnoisy} show comparisons of $\mathbf{x}^*{\ell_1}$ and $\mathbf{x}^*_{BL}$ for denoising piecewise constant, linear, and quadratic functions with one jump with $SNR=0$ dB. Tables \ref{table:1}, \ref{table:2}, and \ref{table:3}, show the error statistics for $SNR=0$ dB as well as other experiments on the same signals at various lower noise levels $SNR=10,20,30$ dB. Bold in these tables indicates the superior performance. There is a significant improvement in accuracy both near edges and in smooth regions.

Finally, note the error bars in these plots and the significance they have with respect to uncertainty quantification. In signal recovery, typically only a single signal estimate is the final result. However, the data collected typically holds more information. In HOTVBL a posterior density is estimated rather than a single point estimate. This allows us to form error bars for the signal of interest as well as its sparse representation. These error bars indicate the certainty of the estimate. They represent the $99\%$ confidence interval associated with the point estimate. These intervals are computed from the diagonal elements of the covariance matrix (i.e., the variance at each point). In addition to potential utility in downstream processing, one general observation these intervals yield is that uncertainty is typically higher in edge regions than in smooth regions.

\begin{table}
\caption{Comparison of relative and maximum errors with varying noise level for piecewise constant function with $m=1$.}
\label{table:1}
\begin{tabular}{lllll}
\hline\noalign{\smallskip}
SNR &  $MaxErr(\bm{x}^*_{BL})$ &  $RelErr(\bm{x}^*_{BL})$ &  $MaxErr(\bm{x}^*_{\ell_1})$ & $RelErr(\bm{x}^*_{\ell_1})$ \\
\noalign{\smallskip}\hline\noalign{\smallskip}
30 dB & \textbf{0.0004} & \textbf{0.0004} & 0.0226 & 0.0108 \\
20 dB & \textbf{0.0005} & \textbf{0.0005} & 0.0968 & 0.0208 \\
10 dB & \textbf{0.0197} & \textbf{0.0189} & 0.1788 & 0.0639 \\
0 dB & \textbf{0.0563} & \textbf{0.0556} & 0.3580 & 0.1710 \\
\noalign{\smallskip}\hline
\end{tabular}
\end{table}

\begin{table}
\caption{Comparison of relative and maximum errors with varying noise level for piecewise linear function with $m=2$.}
\label{table:2}
\begin{tabular}{lllll} 
\hline\noalign{\smallskip}
SNR &  $MaxErr(\bm{x}^*_{BL})$ &  $RelErr(\bm{x}^*_{BL})$ &  $MaxErr(\bm{x}^*_{\ell_1})$ & $RelErr(\bm{x}^*_{\ell_1})$ \\
\noalign{\smallskip}\hline\noalign{\smallskip}
30 dB & \textbf{0.0059} & \textbf{0.0054} & 0.0693 & 0.0164 \\ 
20 dB & \textbf{0.0094} & \textbf{0.0086} & 0.0993 & 0.0516 \\
10 dB & \textbf{0.3863} & \textbf{0.0636} & 0.7247 & 0.1533\\
0 dB & \textbf{0.8681} & \textbf{0.2649} & 0.9273 & 0.3850\\
\noalign{\smallskip}\hline
\end{tabular}
\end{table}

\begin{table}
\caption{Comparison of relative and maximum errors with varying noise level for piecewise quadratic function with $m=3$.}
\label{table:3}
 \begin{tabular}{lllll} 
\hline\noalign{\smallskip}
SNR &  $MaxErr(\bm{x}^*_{BL})$ &  $RelErr(\bm{x}^*_{BL})$ &  $MaxErr(\bm{x}^*_{\ell_1})$ & $RelErr(\bm{x}^*_{\ell_1})$ \\
\noalign{\smallskip}\hline\noalign{\smallskip}
30 dB & \textbf{0.0128} & \textbf{0.0050} & 0.0905 & 0.0227 \\ 
20 dB & \textbf{0.0209} & \textbf{0.0142} & 0.2025 & 0.0630 \\
10 dB & \textbf{0.1226} & \textbf{0.0583} & 0.7387 & 0.1801\\
0 dB & 1.2022 & \textbf{0.3286} & \textbf{1.1926} & 0.3688\\
\noalign{\smallskip}\hline
\end{tabular}
\end{table}

\begin{figure}[t!]
\centering
\includegraphics[width=.49\textwidth]{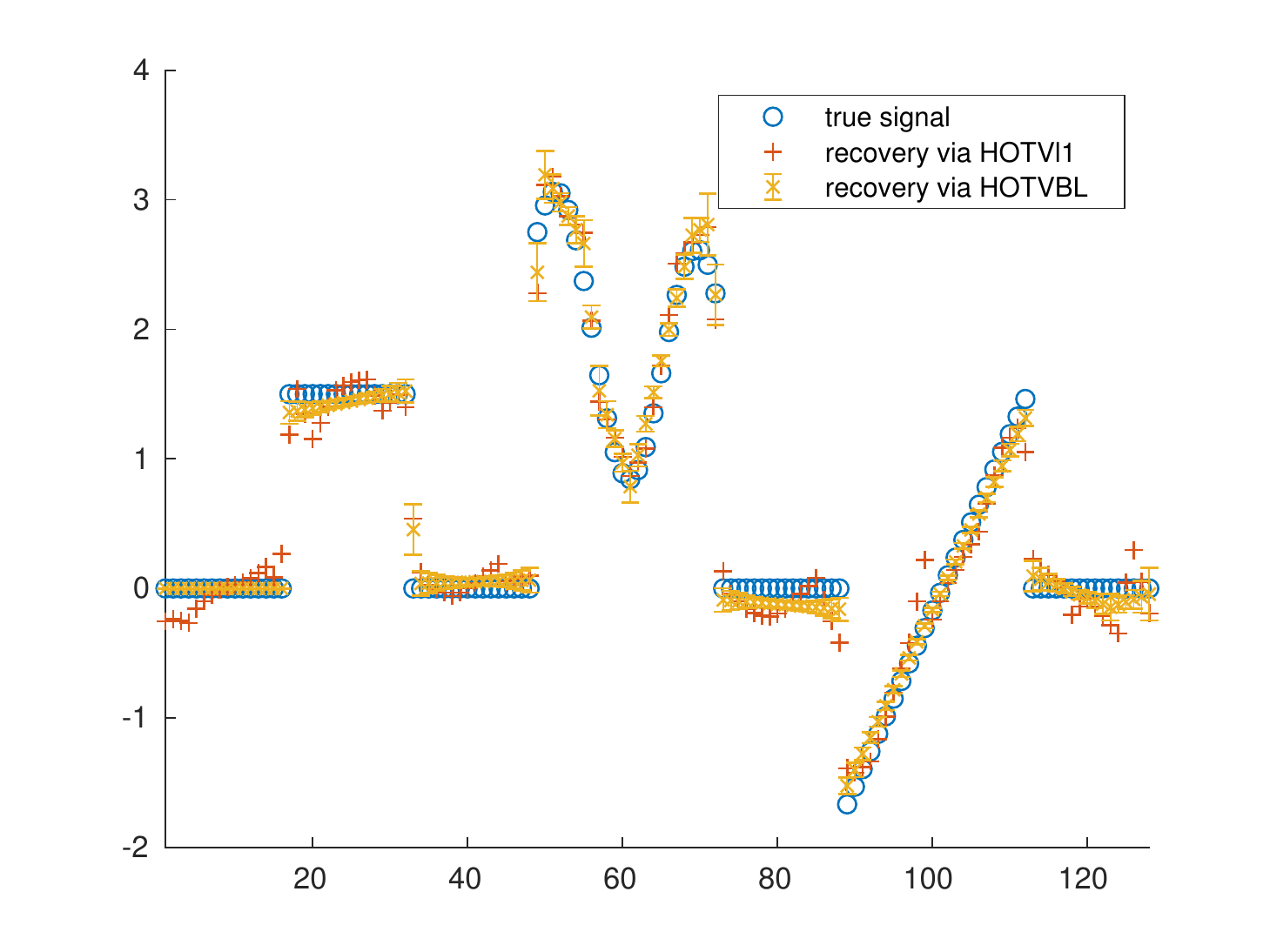}
\includegraphics[width=.49\textwidth]{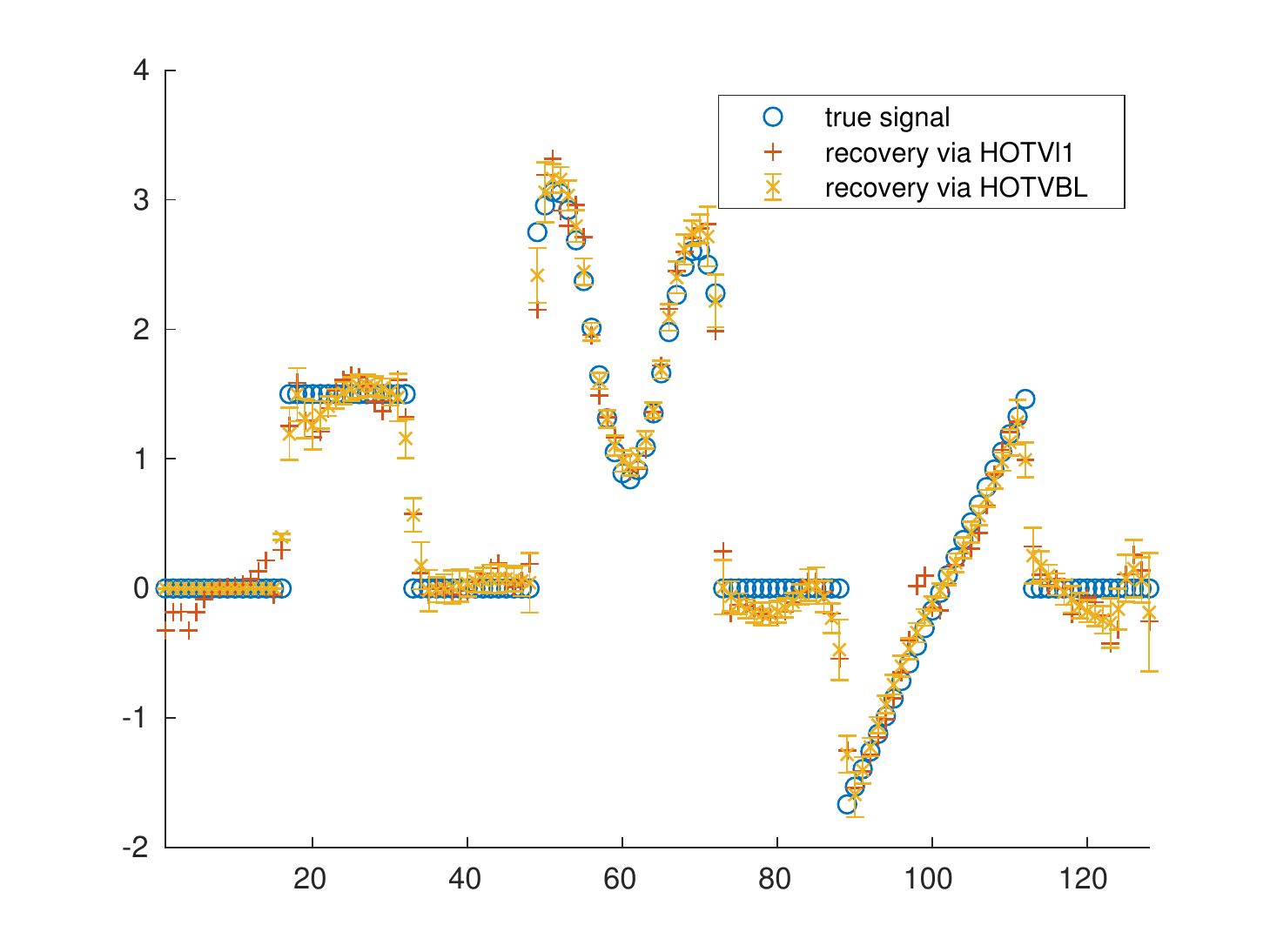}
\caption{Noisy, true, and recovered signals from Fourier data with SNR=10 dB on the left with $m=2$ and on the right with $m=3$.}
\label{fig:fourier10gelb}
\end{figure}

\paragraph{Test 3: Fourier reconstruction of non-ideal function from noisy complex Fourier data.} Complex data can also be used with HOTVBL. E.g., if the signal is real and complex data with complex Gaussian noise is collected, then the model \eqref{eq:model} simply needs to be modified to
\begin{align}\label{eq:complexmodel}
\begin{bmatrix}\mbox{Re}(\bm{y}) \\ \mbox{Im}(\bm{y})\end{bmatrix} &= \begin{bmatrix}\mbox{Re}(\bm{H}_m) \\ \mbox{Im}(\bm{H}_m)\end{bmatrix} \bm{t} + \begin{bmatrix}\mbox{Re}(\bm{n}) \\ \mbox{Im}(\bm{n})\end{bmatrix}.
\end{align}
The problem of reconstructing piecewise smooth signals from spectral or Fourier data, i.e. where $\mathbf{A}$ is the discrete Fourier transform, is a well-studied problem, \cite{gelb2000hybrid,gelb2007reconstruction,gelb2002spectral}. In this problem, discrete Fourier data is collected with SNR = 10 dB.  The Bayesian learning procedure operates exactly as in Section \ref{sec:SBL}. Signals $\mathbf{x}^*_{BL}$ and $\mathbf{x}^*_{\ell_1}$ recovered using $m=2$ and $m=3$ are shown in Figure \ref{fig:fourier10gelb}. In opposition to \emph{Test 2}, the signal used here is piecewise smooth with no $m$ value perfectly sparsifying the signal. The values $m=2$ and $m=3$ were chosen because there are fewer nonzero coefficients in the sparsity representation compared with using $m=1$. In particular, $\bm{\tilde{T}}_1\bm{x}$ had $k=52$, $\bm{\tilde{T}}_2\bm{x}$ had $k=34$, and $\bm{\tilde{T}}_3\bm{x}$ had $k=39$. The maximum and relative errors for $m=1,2,3$ are given in Table \ref{table:4}.

\begin{table}
\caption{Comparison of relative and maximum errors with varying $m$ for piecewise smooth function from Fourier data with SNR=10 dB.}
\label{table:4}
\begin{tabular}{lllll}
\hline\noalign{\smallskip}
$m$ &  $MaxErr(\bm{x}^*_{BL})$ &  $RelErr(\bm{x}^*_{BL})$ &  $MaxErr(\bm{x}^*_{\ell_1})$ & $RelErr(\bm{x}^*_{\ell_1})$ \\
\noalign{\smallskip}\hline\noalign{\smallskip}
1 & 0.5680 & \textbf{0.1262} & \textbf{0.4893} & 0.1329 \\ 
2 & \textbf{0.4542} & \textbf{0.0904} & 0.5396 & 0.1491 \\
3 & \textbf{0.5674} & \textbf{0.1250} & 0.5998 & 0.1670\\
\noalign{\smallskip}\hline
\end{tabular}
\end{table}

\section{Conclusion}\label{sec:conclusion}
This paper presented a Bayesian learning method for inverse problems with an HOTV sparsity prior, including the problem of piecewise smooth function recovery. The standard analysis form HOTV-regularized problem was reformulated by completing the rank of the HOTV analysis operator and inverting it to retrieve an equivalent synthesis operator. This allowed the creation of a Bayesian learning algorithm for piecewise smooth signal recovery that is typically only available for directly sparse problems. Our numerical experiments show that these methods show promise because of their accuracy, the provision of a full posterior density estimate including confidence intervals, and data-driven parameter estimation. HOTVBL is in particular much better suited than the standard HOTV regularized problem in low SNR environments.

Future investigations will include efforts to improve efficiency, perhaps by pre-processing with prior information, which may help to mitigate the cost of implementing HOTVBL for two-dimensional imaging problems. Another potential application for HOTVBL is in effective shock tracking for numerical conservation laws, where the number of grid points are typically much smaller than the number of pixels in a two dimensional image.  HOTVBL may potentially increase the accuracy of the $\ell_1$ regularization techniques for solving conservation laws discussed in \cite{l1,GG2019} for solving numerical conservation laws.  Another benefit in extending the use of HOTVBL to conservation laws is that it will provide a full posterior density estimate as well.

\begin{acknowledgements}
Thank you to Doug Cochran, Aditya Viswanathan, and Theresa Scarnati, for helpful comments and advice on this project.
\end{acknowledgements}

\section*{Conflict of interest}
The authors declare that they have no conflict of interest.

\bibliographystyle{spmpsci}      
\bibliography{refs.bib}   

\end{document}